\documentclass[final,5p,times,twocolumn,number,sort&compress]{elsarticle}
\usepackage[english]{babel}
\usepackage{caption}
\usepackage{fontenc}
\usepackage[pdftex]{hyperref}
\usepackage{subfigure}
\usepackage{amsmath}
\usepackage{amsthm}
\usepackage{amssymb}
\usepackage{psfrag}

\usepackage{color}
\usepackage{soul}

\newlength{\figurewidth}
\usepackage{subeqnarray}
\usepackage[all]{xy}
\usepackage{ctable}
 \usepackage{dcolumn}
\usepackage{array}

\graphicspath{{./figs/}}

\graphicspath{{./FIGURES/}}

\newcommand{\bnabla}{\mbox{\boldmath{$\nabla $}}}
\newcommand{\bv}{\mbox{\boldmath{$v$}}}
\newcommand{\bj}{\mbox{\boldmath{$j$}}}

\begin{document}
\begin{frontmatter}

\title{{\small ACCEPTED MANUSCRIPT} \\      {\small {Published version available at: http://dx.doi.org/10.1016/j.proci.2016.06.066}} \\ -  \\ Propagation  of symmetric and non-symmetric lean hydrogen-air flames in narrow channels: influence of heat losses}
\author{Carmen Jim\'enez}
\ead{Corresponding author: carmen.jimenez@ciemat.es}
\author{Vadim N. Kurdyumov}
\address{Department of Energy, CIEMAT,  Avda. Complutense 40., E-28040 Madrid. Spain.}

\vspace{10em}

\begin{abstract}
   In this paper we present results of direct  numerical simulations of lean hydrogen-air flames freely propagating in a planar narrow channel with varying flow rate, 
using 
detailed chemistry and transport and including  heat losses through the channel walls.
% It is shown that the Poiseuille flow and thermal interaction with the walls together with differential diffusion effects can induce large flame curvatures as well as super-adiabatic temperatures.  Depending on the parameters the resulting curved flame can produce heat at a rate faster than the heat losses, resulting in a curved stationary flame propagating  at a constant speed or can be extinguished by exceedingly fast  heat losses. {Whether the flame survives or extinguishes depends largely on the flame shape, which influences on one hand the flame strength and on the other the temperature gradients at the wall}.
Our simulations show that 
double solutions, symmetric and non-symmetric,  can coexist for a given set of parameters.
The symmetric solutions are calculated imposing symmetric boundary conditions in the channel mid-plane and when this restriction is relaxed non-symmetric solutions can develop.
This indicates that the
 symmetric solutions are unstable to non-symmetric perturbations,
as predicted before within the context of  a thermo-diffusive model and simplified chemistry. It is also found that for lean hydrogen-air mixtures an increase in heat losses leads to 
a discontinuity of the steady state response curve, with  flames extinguishing inside a finite interval of the flow rate. 
%Moreover, we found large differences in the  burning rate of symmetric and non symmetric solutions. 
%Moreover,
 Non-symmetric flames burn more intensely and in consequence are 
much more robust to flame quenching by heat losses to the walls. 
The inclusion of the non-symmetric solutions extends therefore the parametric range for which flames can propagate in the channel. 
%s therefore extended when non symmetric solutions are allowed. 
%The burning rate of the  non-symmetric solutions is  found to be noticeable higher than that of the corresponding symmetric solutions, and therefore the non-symmetric flames are more robust 
% also differ significantly. 
This analysis seems to have received no attention in the literature, even if 
%These results 
it can have important safety implications in 
% the design of
micro-scale combustion devices burning hydrogen in a lean 
premixed way.

\end{abstract}

\vspace{10em}
\end{frontmatter}

%\clearpage

\section{Introduction}

Small-scale combustion has received a lot of interest in recent years as a power and heat generation technology (reviews of the recent technologies and developments can be found in \cite{FernandezPello2002,Maruta2011, KaisareVlachos2012}). This is because it presents advantages over existing small-scale batteries, such as low weight, small size, high power output, fast recharge and long duration. 
%Portable electronic devices demand power generators with less and less weight, longer operating times and faster recharging times. Micro-mechanical devices such as micro-airplanes, micro-rovers, but also pumps and motors, demand high specific energy (low weight, high power output on demand). Burning hydrocarbon fuels or hydrogen in small devices to generate electrical, thermal or mechanical power could be an optimal solution for these technologies.
The obtention of high power in small-scale combustion systems requires, however, completion of the combustion process in a small volume and is then limited by the chemical reaction times. Moreover, the increased surface to volume ratio of small devices results in important heat losses through the walls 
%and can affect 
affecting the stability and even the existence of the flame. In summary, the coupling between fluid dynamics, heat transfer, and chemical kinetics is much more pronounced for these small systems and is a critical element of the design process.

The study of freely propagating flames in ducts is also important
 for safety reasons:  when a  mixture of fuel and oxidizer exists {in the  conducts of a system}, 
it is fundamental to know if a flame can ignite and propagate along it.  After pioneering work in the subject of flames propagating in ducts several years ago \cite{LeeTien1982,LeeTsai1994}, the interest in small-scale combustion systems has revived the studies on the effects of  the channel flow rate \cite{DaouMatalon2001, Tsai2008}, heat-losses \cite{DaouMatalon2002,KurdyumovJimenez2014},  the Lewis number \cite{KurdyumovFernandez2002, Cuietal2004,Cuietal2005,Kurdyumov2011, Galisteoetal2014}
 and thermal expansion \cite{GamezoOran2006,ShortKessler2009,KurdyumovMatalon2013,KurdyumovMatalon2015,JimenezGalisteoKurdyumov2015} on the flame propagation in narrow channels.

%{\color{red} 
%These works show that
%%that 
%%whether a flame survives or extinguishes in a narrow channel with conducting walls 
%%depends largely on the flame shape.
%the channel flow, the thermal interaction with the walls and  preferential diffusion can induce large flame curvatures as well as super-adiabatic temperatures, depending on the parameters.
%Whether the flame survives or extinguishes depends largely on this induced curvature and temperature, which affect the flame burning rate and the temperature gradients at the wall}. 

 Most of these investigations assume from the start that the final flame shape should be symmetric
with respect to a line of symmetry, either axisymmetric in cases
of circular channels or symmetric about the midplane in cases of
two-dimensional channels. However, 
numerical analysis has demonstrated
%it was recently demonstrated 
that  in narrow channels, for mixtures with Lewis numbers smaller than one, steady symmetric and non-symmetric solutions may exist, that the two solutions can co-exist for the same set of parameters, and that when this is the case, the non-symmetric solution is stable while the symmetric solution is usually unstable. For $Le>1$, oscillatory symmetric and non-symmetric solutions can appear.  This was shown for both adiabatic channels \cite{KurdyumovFernandez2002,Kurdyumov2011, Galisteoetal2014, JimenezGalisteoKurdyumov2015} and channels with heat losses through the wall \cite{KurdyumovFernandez2002, KurdyumovJimenez2014}. Of course this is an effect of thermo diffusive instabilities that dominate in narrow channels. In relatively wider ducts a similar effect appears, linked to the hydrodynamic (Darrieus--Landau) instability \cite{Tsai2008, PetchenkoBychkov2005}. Apart from these studies of flames freely propagating in narrow channels, the existence of non-symmetric flames has also been reported in a different configuration, in which flames are stabilized or oscillate in small conducts with heated walls,  both in experimental \cite{Dogwileretal1998} and numerical \cite{Pizzaetal2008a,Pizzaetal2008b,Pizzaetal2009, Miyataetal2015} investigations. It should be noted that the symmetry breaking in that configuration is not necesaarily related to Lewis number or thermal expansion effects, as shown in \cite{Kurdyumovetal2009}, where it was reproduced for $Le=1$ flames under a constant density model.

This effect of symmetry breaking, that should a priori affect, among others,  hydrogen-air flames, was first  demonstrated in the context of a constant density (thermo-diffusive) model and using global one-step Arrhenius chemical kinetics with a single reactant characterized by a single constant Lewis number, so that it could be unmistakebly linked to the diffusive-thermal instability \cite{KurdyumovFernandez2002,Kurdyumov2011, KurdyumovJimenez2014}. Recently, these results have been validated, within a more realistic direct simulation model, including thermal expansion effects and detailed chemical kinetics and transport, for lean  hydrogen-air  flames propagating in adiabatic channels \cite{JimenezGalisteoKurdyumov2015}.
 It was also shown that non-symmetric flames burn much more vigorously than the corresponding symmetric solution, and that consequently the flashback critical conditions predicted by simulations are very different when non-symmetric flames are taken into account.

Here we extend that work \cite{JimenezGalisteoKurdyumov2015} and eliminate the simplifying adiabatic assumption,
incorporating conductive heat losses to the channel walls.   
Our aim is  twofold. First,  we want to  validate  previous results obtained within a simplified model \cite{KurdyumovJimenez2014},  predicting symetric and non-symmetric solutions for $Le<1$ flames propagating in a channel with conducting walls. 
%{\color{red} We wish also to assess whether the bifurcation into symmetric and non-symmetric flames is modified by the effect of heat losses as suggested in \cite{PetchenkoBychkov2005} ??. }  
Secondly, we want to confirm that flame extinction by thermal wall quenching occurs inside a finite interval of channel flow rates, while flames persist  for larger positive or negative flow rates, and obtain a quantitative estimation of the range in which flame quenching occurs for realistic (stable) hydrogen flames.

To this end, we use direct numerical simulations of the Navier-Stokes equations and conservation equations for the species to study 
lean hydrogen-air flames (equivalence ratio $\phi=0.4$, corresponding to flames with effective Lewis number less that one)  in a two dimensional configuration representing an infinite planar channel  of height $1$ mm. We use 8 reactive species and 21 reactions with kinetics given by the San Diego mechanism  \cite{SaxenaWilliams2006}. Note that given the difficulty of obtaining measurements of flames 
%freely
% propagating
  in micro channels, this numerical study will allow for the first time to have quantitative estimations of the parametric range of existence and the propagation speeds (including flashback properties) of  lean hydrogen flames {freely propagating} in small planar conducts {with conducting walls}.

%Given the difficulty of obtaining measurements of flames freely propagating  in micro channels, this numerical study will allow, for the first time:
%\begin{itemize}
%\item To confirm, within a more realistic model, which includes heat expansion effects, previous results \cite{KurdyumovJimenez2014} showing the duplicity of steady solutions, symmetric and non-symmetric for $Le < 1$ flames freely propagating in a narrow channel when heat losses to the wall are present and
%\item To provide a quantitative estimation of the range of existence of symmetric and non-symmetric flames, their propagation speeds and the conditions under which extinction by thermal quenching will occur. Particularly we want to confirm the results found in \cite{KurdyumovJimenez2014} predicting extinction of flames by thermal wall quenching when the reactant flow rate is small and persistence of flames for large positive or negative reactant flow rates.
%%which the thermo-diffusive model can only.
%% As will be shown below, the burning rates of non-symmetric flames and accordingly the critical flow rate below which flashback propagation can be expected are significantly larger for non-symmetric solutions. 
%%\item To assess the influence of  thermal diffusion (the Soret effect) on the flame propagation and the thermal  quenching conditions.
%\end{itemize}
\section{Problem set-up and numerical simulation}

\begin{figure}
\begin{center}
\includegraphics[width=9cm]{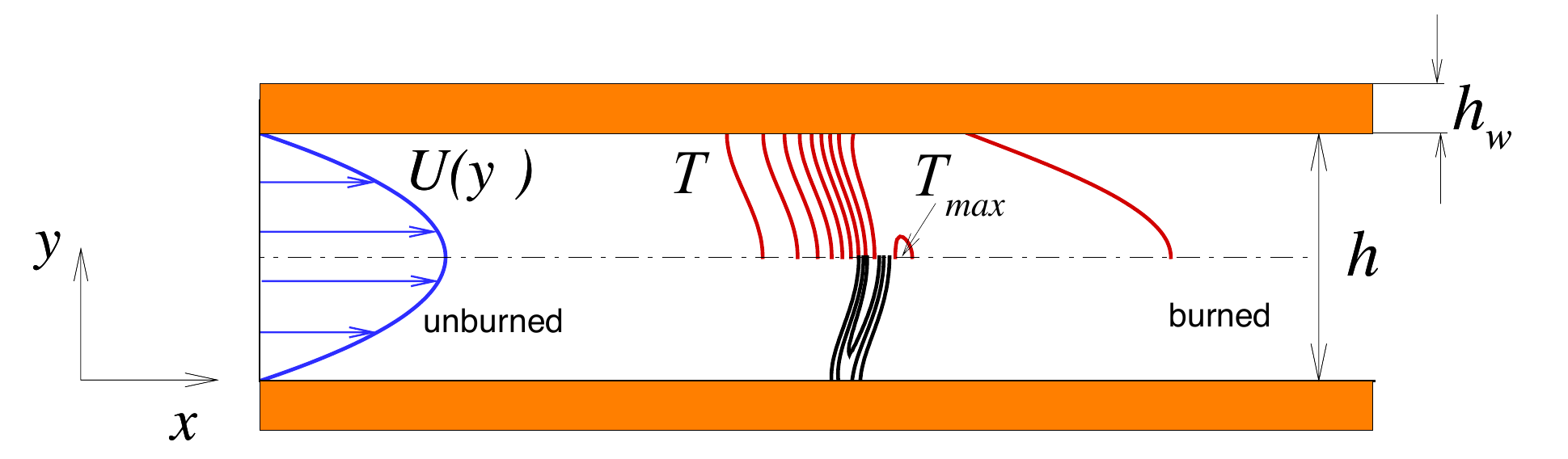}
\end{center}
\vspace{-1em}
\caption{Sketch of the problem.}
\label{fig:sketch}
\end{figure}

Consider a premixed flame propagating in an infinitely long planar channel of width $h$ with walls of width $h_w$. A fuel/air mixture with equivalence ratio $\phi$  at the initial temperature $T_0$ flows through the conduct, driven by a Poiseuille flow with mean velocity $U_0$. If this mixture is ignited, after some transient a curved flame can be established, provided that the heat released in the flame can compensate  the heat losses through the walls. The flame will separate the fresh mixture, far to the left, from the combustion products downstream to the right, as shown in the sketch in Fig.\ref{fig:sketch}. The curvature of the flame is induced by the non-uniform flow and by the heat losses, and enhanced by preferential diffusion if $Le \ne 1$ \cite{KurdyumovFernandez2002, Kurdyumov2011,KurdyumovJimenez2014}.   In the present formulation we are neglecting radiative heat losses and only take into account conductive heat losses to the channel walls, that are considered inert (no surface reactions are included). If these heat losses are sufficiently large the flame can eventually be extinguished (thermal wall quenching). Depending thus on the parameters the flame can be extinguished,  propagate  to the left (flashback), propagate to the right (blowoff), or be stationary in the channel. The specific shape and propagation speed of this flame can only be determined numerically by solving the governing equations of the problem, the conservation equations for mass, momentum, energy and species:
\begin{gather}
\begin{aligned}
\frac{\partial \rho}{\partial t} + \bnabla \cdot \left( {\rho \bv} \right) & =0 ,
\\
\frac{\partial}{\partial t} \left( {\rho \bv} \right) + 
\bnabla \cdot \left( {\rho \bv \bv} \right) & = 
- \bnabla p + \bnabla \cdot \tau,
\\
\frac{\partial}{\partial t} 
\left[ {\rho \left( {e+\frac{v^2}{2}} \right)} \right] + 
\bnabla \cdot \left[ {\rho \left( {e+\frac{v^2}{2}} \right)\bv} \right] & =
 - \bnabla \cdot \left( {p \bv}\right) + 
\bnabla \cdot \left( {\tau \cdot \bv} \right) - \bnabla j_q
\\
\frac{\partial}{\partial t} \left( {\rho Y_i} \right) + 
\bnabla \cdot \left( {\rho Y_i \bv} \right) & = 
- \bnabla \cdot \bj_i + \dot{w}_i ; º: i =1,..,N,
\label{eq1}
\end{aligned}
\end{gather}
where $\rho$, $\bv$, $e$ and $Y_i$ represent, respectively, the density, velocity, internal energy per unit mass and mass fraction of species $i$, $p$ is the pressure, $\tau$ the viscous stress tensor, $\bj_q$ the heat flux, that includes the conductive flux and the diffusive transport of partial enthalpies, $\bj_i$ represents the diffusion flux of species $i$ and $\dot{w}_i$ the mass of species $i$ produced by chemical reactions per unit volume and time.
Body forces and radiation heat fluxes  are assumed to be zero in the present problem. 

To simulate this problem, we solve Eqs.~\eqref{eq1}  
in a long but finite two dimensional domain. We use the compressible solver  NTMIX-CHEMKIN, a high-order accuracy solver designed for the direct numerical simulation of flames with detailed chemistry \cite{NTMIXCHEMKIN}. NTMIX-CHEMKIN features a sixth-order compact differencing scheme \cite{Lele1992} and third-order Runge-Kutta time integration. It incorporates libraries to compute thermodynamic properties and transport coefficients of the species, as well as routines for the calculation of reaction rates based on elemental kinetic mechanisms.  
{{The equation of state for perfect gases, the Navier-Poisson law for the stress tensor and the Fourier law for the conductive heat flux are used, incorporating a mixture averaged model for the viscosity and the thermal conductivity.  Diffusion fluxes  are modeled as proportional to molar fraction gradients  (Hirschfelder-Curtiss model \cite{HirschfelderCurtiss1954})   with a mixture-averaged diffusivity for each species, estimated from the temperature-dependent binary diffusivities.   Note that a correction velocity needs to be incorporated to ensure mass conservation \cite{NTMIXCHEMKIN}.

Species diffusion by temperature gradients (the Soret effect) is included, following \cite{Ibarra1}, \cite{Grcar2009}, where its influence in the onset of cellular instabilities in planar flames, and the shape and heat release rate of curved hydrogen flames was shown. Our previous study  \cite{JimenezGalisteoKurdyumov2015} showed also that for hydrogen flames propagating in adiabatic channels the inclusion of the Soret effect results in appreciably faster flames, by favoring transport of hydrogen to hot regions.  Note that this will require a careful treatment of boundary conditions as explained later at the end of this section. The same studies \cite{Ibarra1, Grcar2009} have shown that the Dufour effect, by which an energy flux is created by species gradients, is negligible. 

As we are interested in the symmetry breaking effect reported for flames with $Le<1$  \cite{KurdyumovFernandez2002, Kurdyumov2011, Galisteoetal2014, KurdyumovJimenez2014},  we have chosen to study  lean hydrogen-air flames with equivalence ratio $\phi=0.4$. Because in this case the limiting reactant is hydrogen this should correspond to an effective Lewis number close to the Lewis number of hydrogen in the mixture, about $Le=0.3$. The recently revisited detailed San Diego chemical kinetics mechanism \cite{SaxenaWilliams2006, SanchezWilliams2014}, that models hydrogen-air combustion using 21 elementary reactions between 8 chemically reactive species, and which has been extensively validated, is used to incorporate the hydrogen combustion chemical kinetics. 

The 
flame propagation  speed of a planar unstrechted hydrogen-air flame of the same equivalence ratio, $U_L$, together with its thermal flame thickness, defined as $\delta_T=  D_{T}/U_L$, where $D_T$ is the thermal diffusivity of the fresh gas mixture, are introduced to specify the dimensionless parameters of the problem. This allows comparison between our simulation results and previous work, usually reported in non-dimensional form. The relevant flame parameters are compiled in Tab.~\ref{tab:parameters}.

\begin{table}[h]
\begin{center}
\begin{tabular}{cccccccc}
\hline
$\phi$ &	 $U_{L }$ (cm/s) & $\delta_T$  (mm) 	 &  $h$ (mm) &     $d$ & $m$	& $b$\\	
\hline 	
   0.4 & 21	& 0.159	 &  1 & 40  & [-6:8] & [0:5] \\
\end{tabular}	
  \caption{Simulation parameters. { Note that $U_L$ and $\delta_T$ correspond to the computed speed and thickness of the planar unstretched hydrogen flame when the Soret effect is included.}}
 \label{tab:parameters}
  \end{center}
  \end{table}

The computational domain measures $1$ mm in the y-direction, 
 while the size in the x-direction is typically $L=10$ mm, extended to $15$ mm or even $20$ mm for some cases where  a longer domain is required to accommodate a highly curved flame. The domain is discretized on a uniform rectangular grid containing typically $601 \times 61$ nodes for a channel measuring $10 \: \mbox{mm} \times 1 \mbox{ mm}$. Grid refinement studies have shown no appreciable difference in the  obtained results when the grid resolution is halved.  

Boundary conditions at the left boundary ($x=0$) are imposed as a Poiseuille velocity profile driving a mixture of hydrogen-air at temperature $T_0 =300$ K,  initial pressure $P_0= 1$ atm and equivalence ratio $\phi=0.4$.
At the right boundary ($x=L$), partially non-reflecting boundary conditions are imposed using the NSCBC methodology  \cite{PoinsotLele1992,Baum1995}. These conditions allow to specify a pressure level at the far field and at the same time permit physical or numerical waves to leave the domain. 

Boundary conditions at the wall represent the gas-solid interaction. If the external surface of the wall is asumed to be at the fresh gases temperature, $T_0$,  and the wall thickness $h_w$ is assumed to be small, $h_w/h \ll 1$, then the temperature distribution within the wall can be taken to be linear  \cite{KurdyumovJimenez2014}:
\begin{gather}
\begin{aligned}
T_B & = T_{y=0} + (T_{y=0}-T_0) \: y/h_w \\
T_T & = T_{y=h} + (T_0-T_{y=h}) \: (y-h)/h_w ,
\end{aligned}
\end{gather}
inside the bottom and top wall, respectively.

Imposing continuity of the heat flux in the gas and the internal surface of the solid wall results in a relation between the temperature gradient in the gas and solid side of the boundary
which allows to write boundary conditions for the gas temperature gradient at the wall \cite{KurdyumovJimenez2014}:
\begin{gather}
\begin{aligned}
\left.\frac{\partial T}{\partial y}\right|_{y=0} & =  \frac{\lambda_w}{\lambda_g h_w} (T_{y=0}-T_0) \\
\left.\frac{\partial T}{\partial y}\right|_{y=h} & = \frac{\lambda_w}{\lambda_g h_w} (T_0-T_{y=h}),
\end{aligned}
\end{gather}
where $\lambda_g$ and $\lambda_w$ are the thermal conductivity of the gas and the solid surface respectively.
These conditions  together with no-flux condition for the species and no-slip velocity condition represent the gas-wall interaction.

When the species diffusion fluxes are modelled by a  mixture-based diffusivity,  the no-flux condition at the walls translates simply to a zero gradient condition for the molar fractions of each species.
However, if
 the thermal diffusion (Soret effect) is included, 
the no flux condition imposes that the combined flux created by molecular diffusion (dependent on the gradients of molar fractions) and thermal diffusion (dependent on the temperature gradient)  should be zero for each species \cite{Markatou1991,PoppSmookeBaum1996}. While for adiabatic walls  this boundary condition reduces to  a zero gradient condition on the molar fraction of each species, this is not true for the general case of walls with heat losses.

The heat transfer parameter 
 $ \frac{\lambda_w}{\lambda_g h_w}$ depends of the wall thickness and thermal conductivity, which are taken to be constant, and the gas conductivity, which varies with temperature in the flame.
We will use as reference a non-dimensional parameter given by the  value of the heat transfer parameter in the unburned gases  scaled with the flame thickness, similar to that introduced in \cite{KurdyumovJimenez2014}:
 \begin{equation}
 b=  \frac{\lambda_w}{\lambda_g^0}\frac{\delta_T}{h_w},
\label{eq:b}
 \end{equation}
where $\lambda_g^0$ is the thermal conductivity of the unburned mixture.
 $b=0$ corresponds to adiabatic walls while the limit $b=\infty$ represents isothermal walls.

Following \cite{Kurdyumov2011} the heat transfer parameter $b$, together with the  dimensionless mean flow rate in the channel, scaled by the laminar flame speed:
\begin{equation}
                                m=U_0/U_L,                                                              
\label{eq:m}
\end{equation}
and the Damk\"ohler number:
\begin{equation}
                               d=(h/\delta_T )^2,
 \label{eq:dam}
 \end{equation}
  will be taken as the dimensionless parameters of the problem. A  channel gap   $h= 1$ mm is used in the present study, corresponding to a fixed Damk\"ohler number  $d=40$, similar to the value studied in \cite{KurdyumovJimenez2014,JimenezGalisteoKurdyumov2015}.The flow rate $m$ is varied between $m=-6$ and  $m=8$ and the heat transfer parameter $b$  between $0$ and $5$. These channel-related parameters are also reported in Tab.~\ref{tab:parameters}.

The  simulations are initialized with a planar lean ($\phi=0.4$) hydrogen-air flame, located at a mid-position in the x-axis. This flame solution is superimposed to an initial Poiseuille flow with mean flow rate $m$.  A negative $m$ corresponds to a fresh reactant mixture flowing to the left in the sketch of Fig.~\ref{fig:sketch}, away from the flame. A positive value of $m$ corresponds to fresh gases flowing towards the flame.
Since both symmetric and non-symmetric solutions for the flame are expected, and typically when both type of solutions exist for the same set of parameters the symmetric solution is  unstable \cite{Kurdyumov2011, Galisteoetal2014, KurdyumovJimenez2014}, two kinds of unsteady calculations are undertaken for every value of  $m$ and $b$: 
\begin{itemize}
\item In the first computation, the full domain is included in the simulation, and a small non-symmetric perturbation in the shape of a hot spot  just upstream of the flame at a location  $y=3/4 \:  h$ is added to the initial conditions. In this calculations, the unsteady simulation will converge to the steady stable solution, be it symmetric or  non-symmetric. {Note that the non-symmetric hot spot is only introduced to accelerate the transition. When the symmetric flame is unstable, even without this initial perturbation, the solution eventually converges to one of the two possible non-symmetric flames.}
\item In the second computation the possibly unstable symmetric solution is sought. To this end, the computational domain is reduced to a half channel ($0 \le y\le h/2$) and symmetric boundary conditions are imposed at the channel axis $y=h/2$. In such a way the symmetric solution is forced, even when it is not stable.
\end{itemize}

A symmetry factor defined as \cite{Kurdyumovetal2009}:
 \begin{equation}
 {{S}}= \frac{1}{\left(T_{ad}-T_0\right) h^2 } \int_{-\infty}^{\infty} \int_{0}^{h/2} \left[ T(x,y)-T(x,h-y) \right] {\rm d}x \: {\rm d}y,
 \end{equation}
 equal to zero for symmetric flames, shall be used to distinguish symmetric and non-symmetric solutions.

After initialization, the flame will change its curvature, depending on the flow rate $m$ and the heat transfer parameter $b$. The shape it acquires will determine the wall temperature gradient and therefore the heat loss rate. 
The balance between the burning rate and the heat losses will determine whether a flame can propagate for a given parameter set. Note that for large values of the heat transfer parameter $b$ extinction could occur in the transient before the flame has adopted its final curvature (and its final burning rate). For this reason, for large values of $b$   initial conditions corresponding to the steady flame solution in an adiabatic channel were used. In this way the  large initial curvature (and therefore burning rate) prevents the occurrence of this {spurious} extinction.
 
In the case were a flame can self sustain, its propagation speed will depend on the burning rate and the reactants flow rate. We name the flame propagation speed with respect to the wall  $U_f$ in dimensional units or $u_f=  U_f/U_L$  when scaled with the laminar flame speed, and define it as positive when the flame propagates towards the left (flashback).

The consumption speed or burning rate (scaled with the laminar flame speed) can be measured by evaluating:
\begin{equation}
u_c= - \frac{1}{\rho_0 Y_0^F h \: U_L }  \int_{-\infty}^{\infty} \int_{0}^{h}   \dot{\omega}_F  \: {\rm d}x \: {\rm d}y,        
\label{eq:sc}         
\end{equation}
where $Y_0^F$ and $\rho_0$ are the fuel mass fraction and the density in the fresh gases and  $\dot{\omega}_F$  is the volumetric mass fuel consumption rate, defined negative, hence the minus sign ensures a positive value for $u_c$.  For a steadily propagating flame, moving as a rigid structure without shape changes, the consumption speed is equal to the flame propagation speed with respect to the fresh gases.

The dimensionless flame propagation speed with respect to the wall can then be estimated as:
\begin{equation}
                                       u_f= u_c-m,      
\label{eq:uf}                                    
\end{equation}                   
which is positive if consumption is faster than the reactant mixture inflow rate, in which case the flame propagates towards the left, and negative when the reactants flow is faster than consumption, in which case the flame propagates towards the right. When the consumption speed equals the inflow rate the flame consumes the reactants at the same pace they are fed to the reaction zone, and is therefore stationary in the channel.

In order to keep the propagating flame inside the computational domain when $u_f \ne 0$, we adopt a reference frame moving with the flame as in \cite{JimenezGalisteoKurdyumov2015}.
This is implemented by {estimating at every time step 
the propagation speed $u_f$ via the expressions in Eqs.~\eqref{eq:sc} and \eqref{eq:uf},  
and 
substracting $u_f$  from the inlet and wall boundary conditions for the flow velocity.
 In all the  cases considered in this work the final flame is  stationary  in the reference frame moving at the final propagation speed  $u_f$}.

\section{Results and discussion}

\subsection{Symmetric and non-symmetric steady solutions.}

Let's first focus on the time evolution  of two simulations of flames propagating in the channel with reactant flow rates $m=-6$ and $m=+2$ and heat transfer parameter $b=0.05$, as  presented in Fig.~\ref{fig:timehistory}. Both simulations are computed in the full computational domain, with no assumptions on the symmetry of the solution. The two unsteady calculations  converge to steady solutions with different burning rates and symmetry properties. The final steady solution is symmetric with $u_{\rm{c}}=4.6$ for $m=-6$,  and non-symmetric with $u_{\rm{c}} = 5.6$ for $m=2$. Given that these are unsteady simulations, 
the final solutions are necessarily stable. This confirms that, as was the case in lean hydrogen flames in adiabatic channels \cite{JimenezGalisteoKurdyumov2015},
depending on the flow rate parameter, symmetric and non-symmetric steady stable solutions can exist.
 It should be mentioned that when a non-symmetric flame exists, 
the corresponding symmetric solution can also be found using a half-domain computation, as will be shown later.

\begin{figure}
 \begin{center}
 \includegraphics[width=9cm]{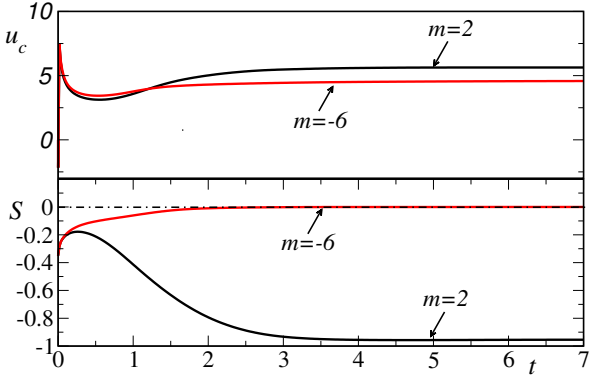}
 \end{center}
\vspace{-2em}
 \caption{The time history of the flame burning rate $u_{\rm c}$ (top) and the symmetry indicator ${S}$ (bottom)  for two values of the flow rate $m$ in the simulation
 of flames propagating in a  channel with $b=0.05$.}
  \label{fig:timehistory}
\end{figure}

Even if the flames presented in Fig.~\ref{fig:timehistory} have similar burning rates, because of their different shapes they are affected differently by an increase in heat losses. Indeed, we will show that  the fastest burning flame ($m=+2$) is extinguished for a smaller $b$ value than the weakest flame ($m=-6$).
Figures \ref{fig:one} and \ref{fig:two} present the final steady state flames for different values of the parameters. In these figures we plot contours of the
 reduced temperature, 
                               $\theta=(T-T_0 )/(T_{ad}- T_0 )$,                               
and the reduced volumetric heat release rate,
                                 $ Q'=  Q/Q^{max}_L$,
where $T_{ad}$ and  $Q^{max}_L$   are the adiabatic flame temperature and the maximum volumetric heat release rate value in the initial planar flame.

%
%two column version
\begin{figure*}
\begin{center}
{\scriptsize a}
 \includegraphics[width=9.cm]{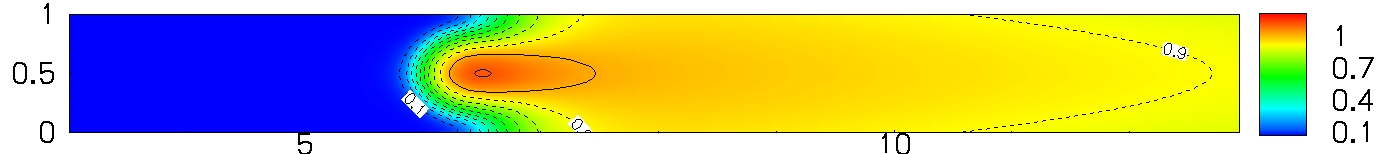}  
 \includegraphics[width=9.cm]{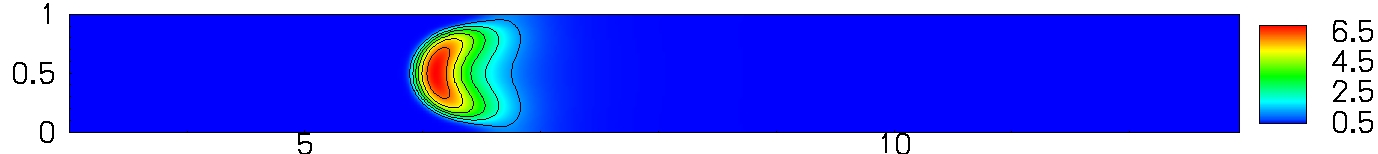}

{\scriptsize b}  \includegraphics[width=9cm]{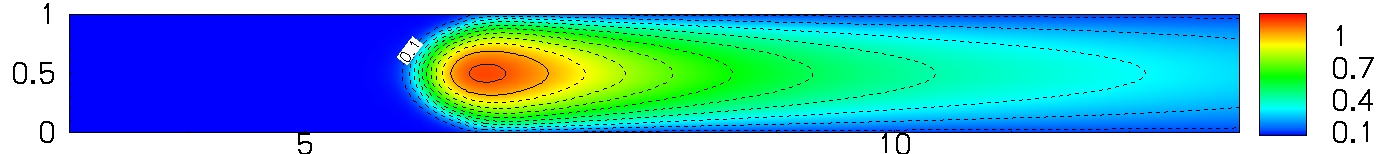}  
\includegraphics[width=9cm]{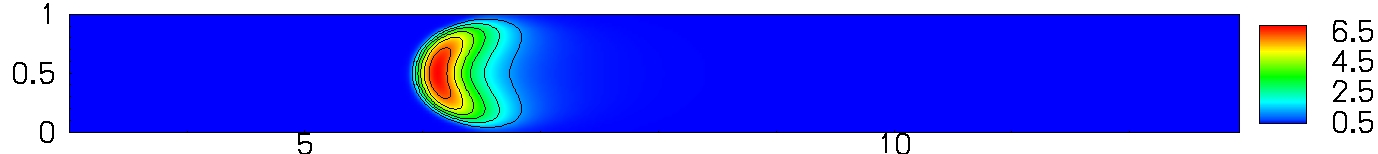}
 \end{center}
\vspace{-1em}
 \caption{The reduced temperature $\theta$ (left) and the heat release rate $Q'$  (right) of the steady flame solution obtained for  $m=-6$ and $b=0.05$ (a) and $b=5$ (b).  The temperature isocontours are plotted at $0.1$ intervals and marked with solid lines for $\theta \ge 1$. The heat release isocontours are plotted at $\Delta Q'=1$ intervals.}
 \label{fig:one}

\begin{center}
 {\scriptsize a}  \includegraphics[width=9cm]{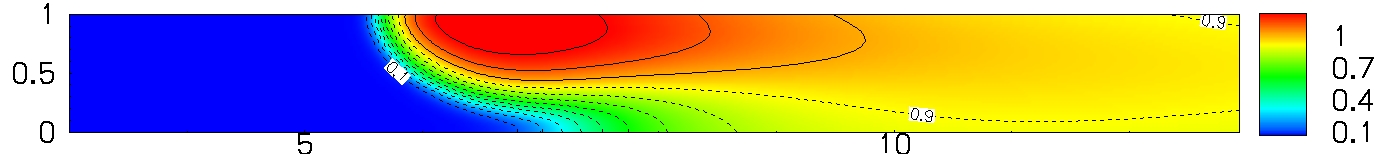}  
\includegraphics[width=9cm]{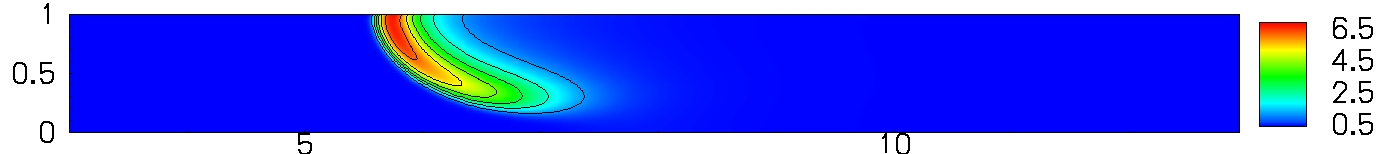}

{\scriptsize b}   \includegraphics[width=9cm]{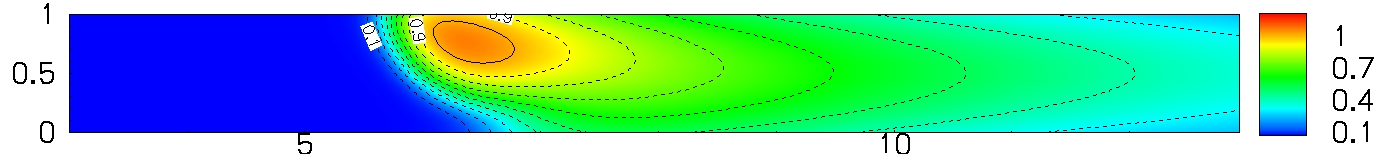}  
\includegraphics[width=9cm]{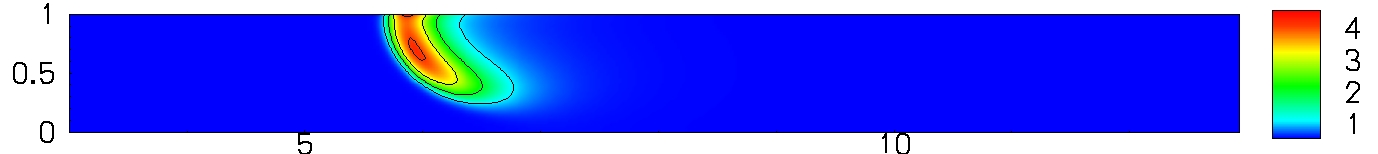}
 \end{center}
\vspace{-1em}
 \caption{The reduced temperature $\theta$ (left) and the heat release rate $Q'$  (right) of the steady flame solution obtained for  $m=+2$ and $b=0.05$ (a) and $b=0.35$ (b).  The temperature isocontours are plotted at $0.1$ intervals and marked with solid lines for $\theta \ge 1$. The heat release isocontours are plotted at $\Delta Q'=1$ intervals.}
 \label{fig:two}

 \begin{center}
{\scriptsize a}  \includegraphics[width=9cm]{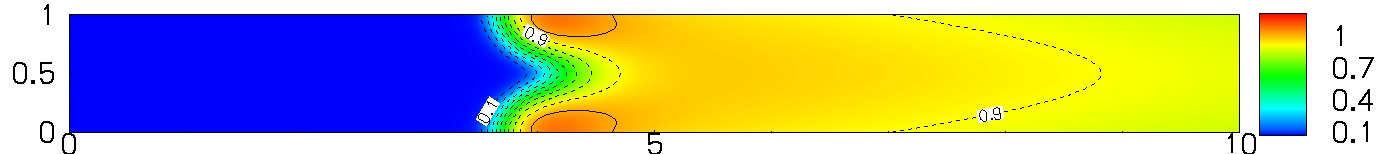}  
\includegraphics[width=9cm]{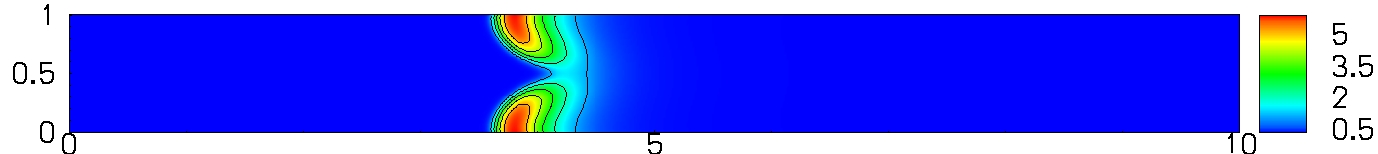}

{\scriptsize b}  \includegraphics[width=9cm]{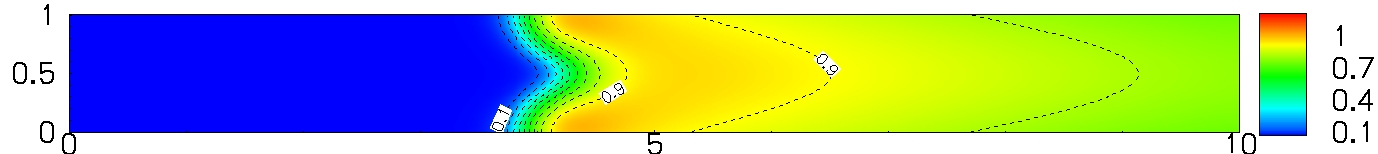} 
\includegraphics[width=9cm]{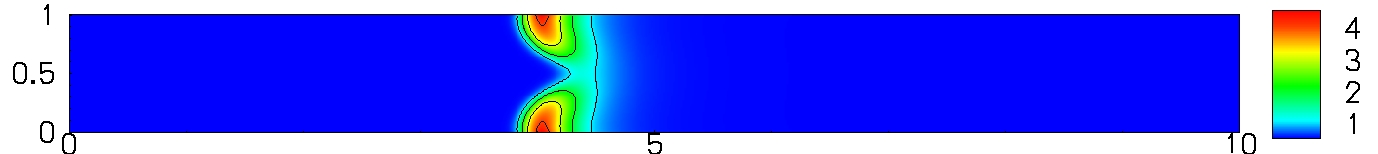}
 \end{center}
\vspace{-1em}
 \caption{The reduced temperature $\theta$ (left) and the heat release rate $Q'$ (right) of the steady flame solution obtained in a half channel imposing symmetry about the mid $y$-axis for $m+=2$ and $b=0.05$ (a) and $b=0.075$ (b).  The temperature isocontours are plotted at $0.1$ intervals and marked with solid lines for $\theta \ge 1$. The heat release isocontours are plotted at $\Delta Q'=1$ intervals.}
 \label{fig:three}
\end{figure*}

Figure \ref{fig:one} a corresponds to the case where reactants flow to the left, accompanying the flame, with  $m=-6$ and with heat transfer $b=0.05$. 
For this flow rate the resulting flame is symmetric, as shown by the $S=0$ value of Fig.~\ref{fig:timehistory}, and presents maxima for the heat release rate and temperature in the middle of the channel. This results in small wall temperature gradients and therefore small  heat losses. The flame still burns  when the heat transfer parameter is increased by a factor $100$,   as shown in Fig.~\ref{fig:one} b. The burning rate of the solution with $b=5$  is only  slightly smaller than that of the flame with $b=0.05$ ($u_c=4.5$ for $b=5$ versus $u_c=4.6$ for $b=0.05$). Thus this flame is, because of its   shape, very robust to wall quenching.

Figure \ref{fig:two} a corresponds to the case with $m=+2$  and $b=0.05$.  In this case the flow is directed to the right, opposed to the flame. The resulting flame is  clearly non-symmetric, as corresponds to the non-zero $S$ value of Fig.~\ref{fig:timehistory}, and presents a large curvature, which results in a flame burning intensely ($u_{\rm{c}} = 5.6$). The maximum temperature and heat release rate are located in the vicinity of the wall, resulting in large wall temperature gradients, that can induce large heat losses. When the heat transfer parameter is small, $b=0.05$, the flame presents still high temperatures, exceeding $1.2 \: T_{ad}$. As $b$ is increased heat losses grow rapidly, and for $b=0.35$ the flame is close to extinction with a maximum temperature only slightly over the adiabatic temperature $T_{ad}$ as pictured in Fig.~\ref{fig:two} b.

Figure \ref{fig:three} a presents a flame computed in a half channel, with imposed symmetry about the mid $y$-axis, for the same set of parameters  ($m=+2$ and $b=0.05$). This flame presents a   shape concave towards the reactants, with maximum heat release and temperature close to the walls. The curvature is smaller  than that of the non-symmetric flame obtained in the full domain for the same parameters and the maximum temperature is only slightly over $T_{ad}$. As a consequence the burning rate is relatively weak ($u_c=1.8)$ while  wall temperature gradients are important. The result is that for a heat transfer as small as $b=0.075$ the symmetric solution is very close to extinction (Fig.~\ref{fig:three} b). Even if no stability analysis is done in this work,  we can conjecture that these solutions are unstable, because when the symmetry condition is {relaxed}, the solution found  is the non-symmetric flame, as shown in 
Fig.~\ref{fig:timehistory}.  This is a clear example of a case in which imposing symmetry of the flame would lead to erroneous predictions, both in terms of the intensity of the flame burning and of its wall quenching behavior.

\subsection{Flame extinction for symmetric and non-symmetric simulations}

We will now explore the behavior of flames in the full range of variation of $m$ for several values of the heat transfer parameter $b$. Figure \ref{fig:b05} shows a plot of the computed flame
 propagation speed $u_f$ as a function of the non-dimensional flow rate $m$ for adiabatic channels ($b=0$, open squares) and channels with weakly conducting walls ($b=0.05$, filled circles). 
Solid lines correspond to full domain calculations and dashed lines to calculations in a half channel with imposed symmetry about the mid $y$-axis. 
One can see that for large negative  values of $m$
all the curves are close to each other. Indeed, only symmetric solutions exist for $m<-3$ and the results of full and half domain computations are identical, so that the dashed and solid curves coincide for each value of $b$.  

For flames computed in the full domain (solid lines), the curve corresponding to $b=0.05$ stands at a short distance below the adiabatic flames curve ($b=0$).  The effect of weakly conducting walls is in this case to lower the burning and propagating flame speeds, but heat losses are not sufficiently strong to produce extinction.

However, for simulations in the half channel with imposed symmetry (dashed lines), there is a large interval between about $m\approx-3$ and $m\approx2$ where flames can not sustain heat losses and are extinguised even for this low value of $b$.  By increasing the flow rate a branch or burning solutions is found for $m\ge 2$. This behavior,  predicting flames extinguising in a finite interval of values of $m$  and burning again for larger values could seem paradoxical at first sight. Nevertheless, it is easily explained by the larger curvatures imposed in the flames by stronger convective forcing, which result in more intensely burning flames. 

 In the range of relatively slow flow between  $m=-3$ and $m=2$ the symmetric solutions are nearly planar. This small curvature implies slow burning rates, so that extinction occurs even when the heat transfer to the wall is weak. As the flow rate is increased over $m=2$ the flame curvature increases and so do the burning and propagation speeds (remark that for $m$ just above $2$ there is a sharp increase in the propagation speed for the adiabatic symmetric flames of Fig.~\ref{fig:b05}). These more intense flames are more resistent to quenching by heat losses, and therefore a branch of burning solutions appears for $m > 2$.

As the heat transfer parameter is increased to  $b=0.1$, the interval where  flames are extinguised also increases. Figure \ref{fig:b1} presents a comparison of the propagation speed $u_f$ in adiabatic channels and channels with walls with $b=0.1$, computed in the full domain (solid lines) and the half channel domain (dashed lines). An interval of symmetric solutions (circles with dashed lines)  between $m\approx-4.5$ and $m\approx2.5$ corresponds now to extinguished flames. A smaller gap appears where the non-symmetric flames are also extinguished. 

Further increases in the heat transfer parameter keep the tendency of increasing the interval where flames are extinguished. As can be seen in Fig.~{\ref{fig:b235}, for $b=0.2$ no flame can exist between the flow rates of  $m\approx-4.5$ and  $m\approx-1$, for $b=0.3$ the flame quenching interval is extended up to flows with $m\approx1$ and for $b=0.5$ up  to flows with $m \approx 5.5$.  Only steady state results of simulations computed in the full domain, corresponding to stable and therefore realizable  solutions are included in this figure. 

Note that symmetric, half-domain simulations would have predicted  a larger flame quenching interval for each of the values of  $b$. 
Note also that the predicted burning and propagation rates would be significantly lower.  This means that calculations
 assuming symmetry about the channel mid-axis would underpredict the parametric range where flames can propagate along a given channel
as well as the range where flashback propagation can occur. Evidently this underprediction could result in important safety issues if these calculations are used in the design or characterization of small combustion systems.

%two column version
\begin{figure}
 \begin{center}
   \includegraphics[width=9cm]{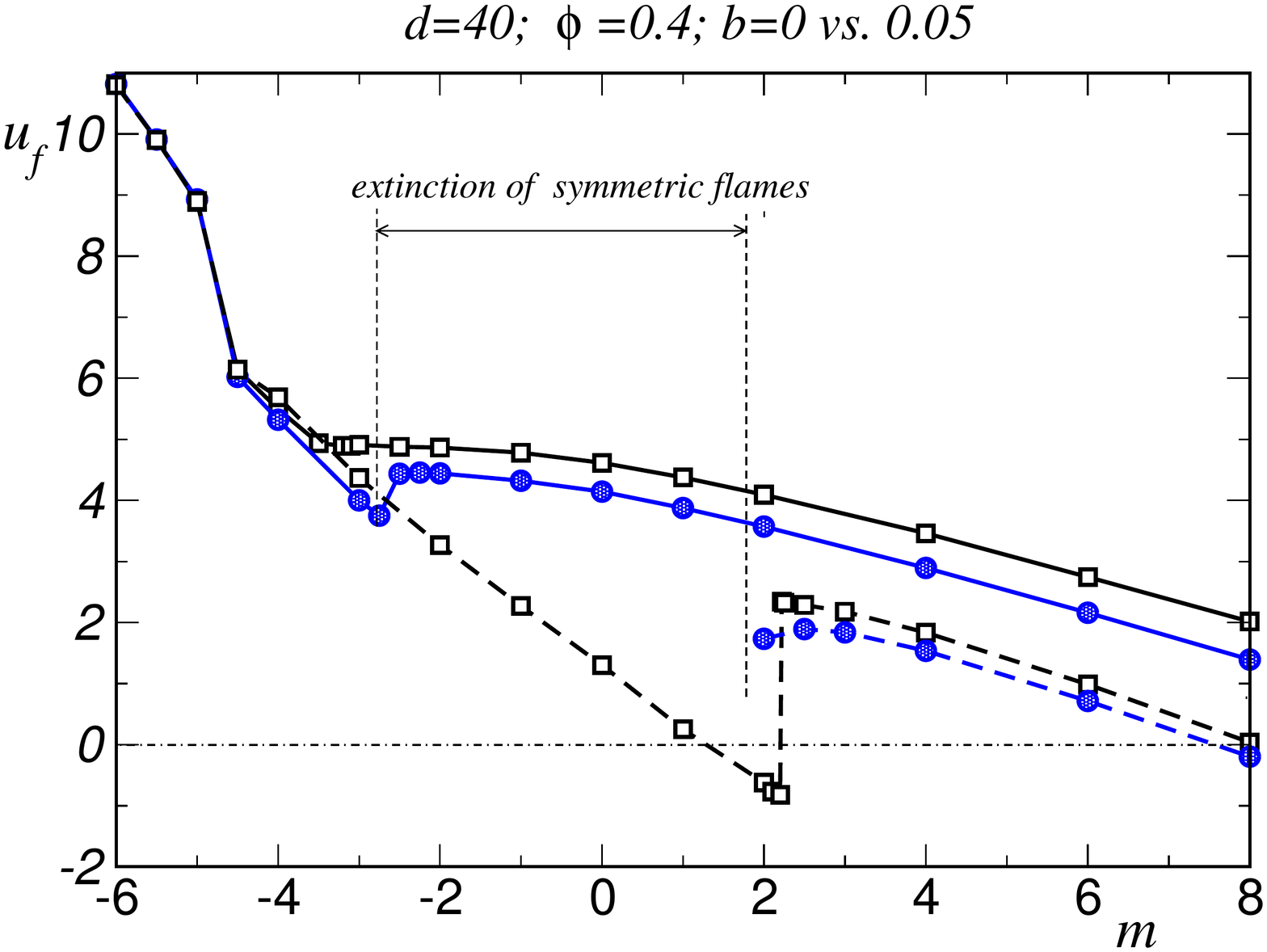}
 \end{center}
\vspace{-2em}
 \caption{The propagation speed $u_f$ of non-symmetric (solid lines) and symmetric (dashed lines) flame computations  as a function of the flow rate $m$ for $b=0$ (empty squares) and $b=0.05$ (filled circles).}
  \label{fig:b05}
 \begin{center}
   \includegraphics[width=9cm]{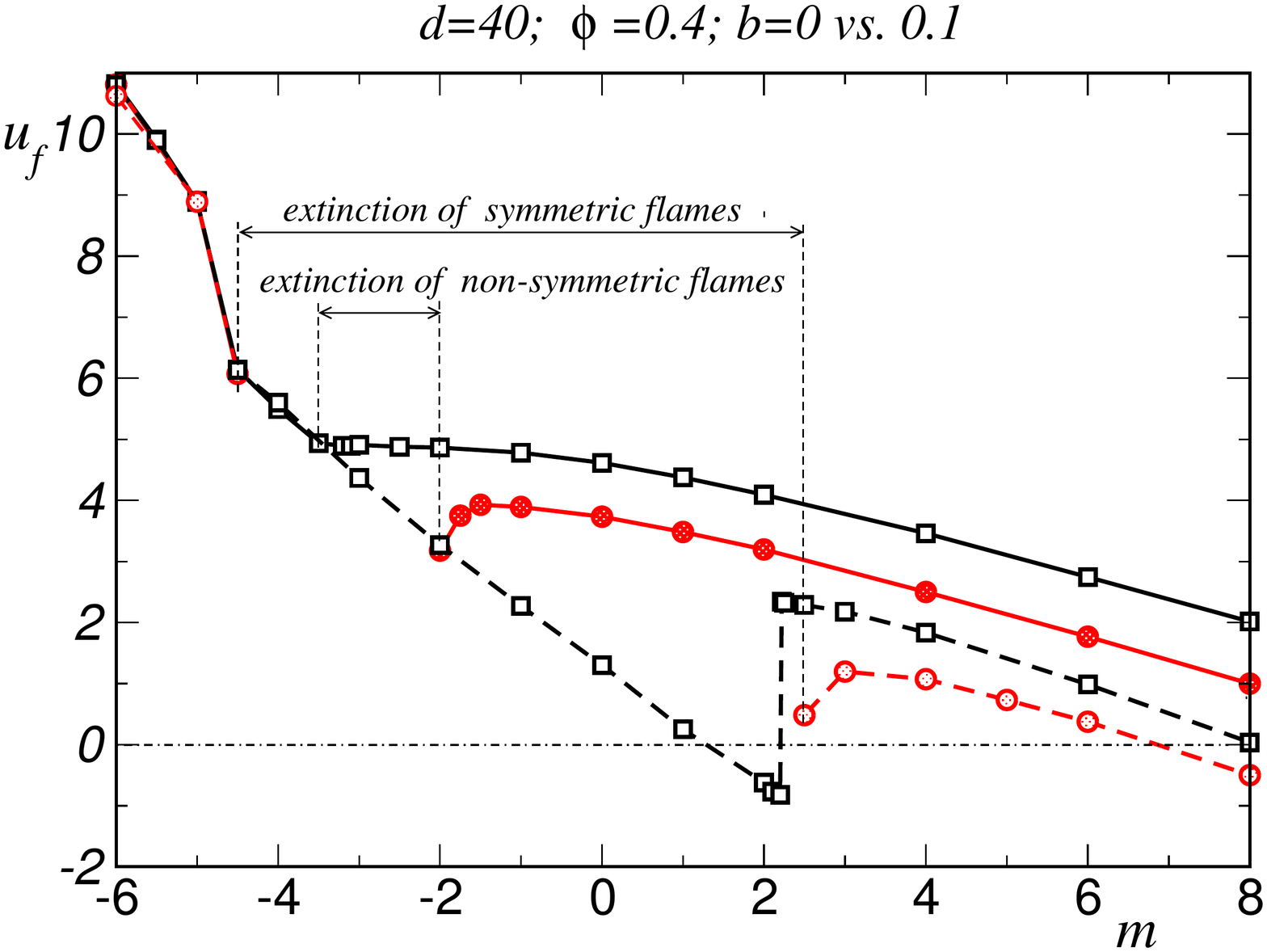}
 \end{center}
\vspace{-2em}
 \caption{The propagation speed $u_f$ of non-symmetric (solid lines) and symmetric (dashed lines) flame computations  as a function of the flow rate $m$ for $b=0$ (empty squares) and $b=0.1$ (filled circles).}
  \label{fig:b1}
 \begin{center}
   \includegraphics[width=9cm]{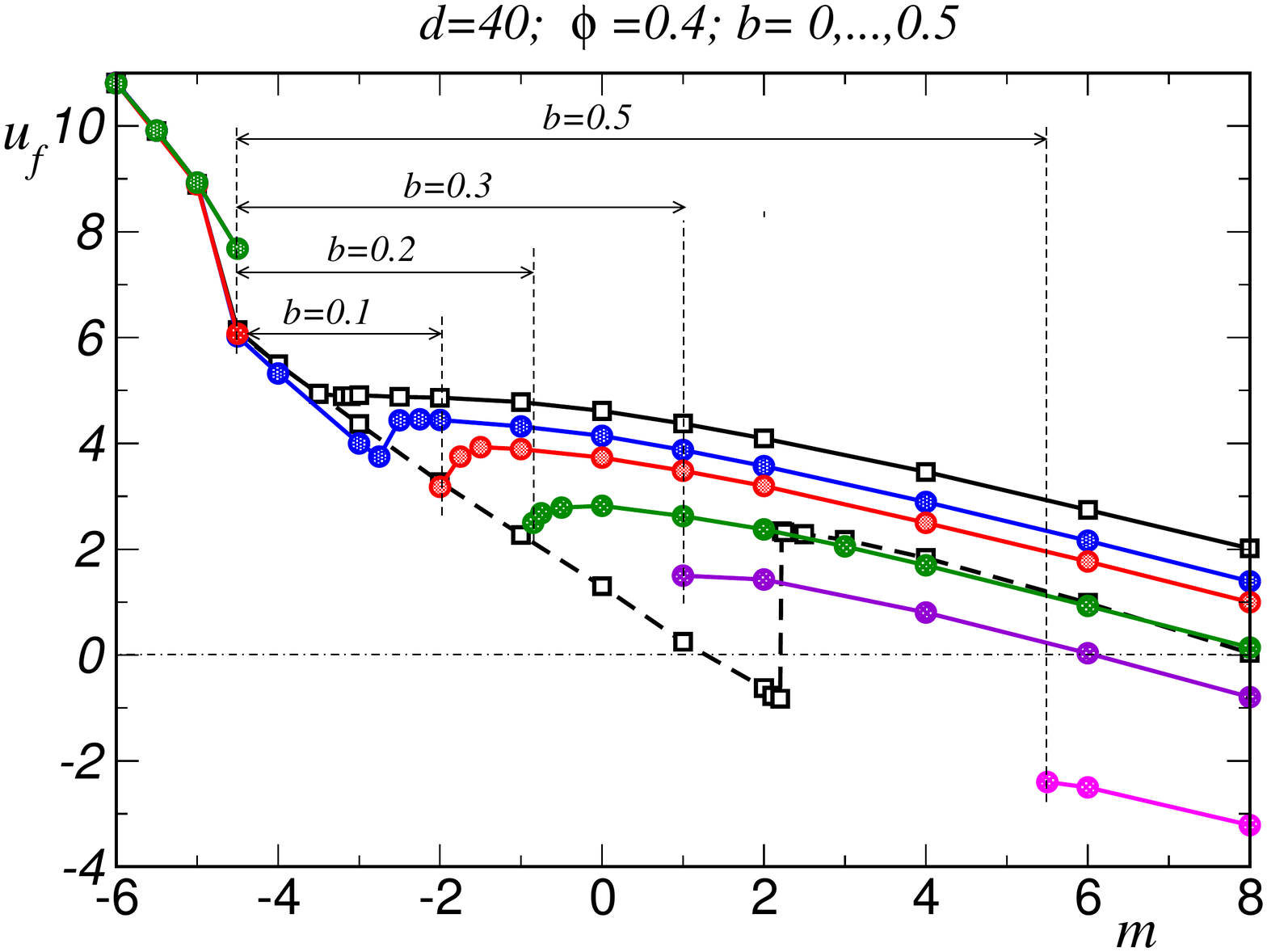}
 \end{center}
\vspace{-2em}
\caption{The propagation speed $u_f$ of steady stable flames  as a function of the flow rate $m$ for $b=0$ (empty squares) and $b=0.05, 0.1, 0.2, 0.3$ and $0.5$ (filled circles). The propagation speed obtained for adiabatic symmetric domain computations is also represented with empty squares and dashed lines. }
 \label{fig:b235}
\end{figure}
%
%%one column version
%\begin{figure}
% \begin{center}
%   \includegraphics[width=9cm]{fig6}
% \end{center}
%\vspace{-2em}
%\caption{The propagation speed $u_f$ of non-symmetric (solid lines) and symmetric (dashed lines) flame computations  as a function of the flow rate $m$ for $b=0$ (empty squares) and $b=0.05$ (filled circles).}
%  \label{fig:b05}
%\end{figure}
%
%
%\begin{figure}
% \begin{center}
%   \includegraphics[width=9cm]{fig7}
% \end{center}
%\vspace{-2em}
% \caption{The propagation speed $u_f$ of non-symmetric (solid lines) and symmetric (dashed lines) flame computations  as a function of the flow rate $m$ for $b=0$ (empty squares) and $b=0.1$ (filled circles).}
%  \label{fig:b1}
%\end{figure}
%
%
%\begin{figure}
% \begin{center}
%   \includegraphics[width=9cm]{fig8}
% \end{center}
%\vspace{-2em}
% \caption{The propagation speed $u_f$ of steady stable flames  as a function of the flow rate $m$ for $b=0$ (empty squares) and $b=0.05, 0.1, 0.2, 0.3$ and $0.5$ (filled circles). The propagation speed obtained for adiabatic symmetric domain computations is also represented with empty squares and dashed lines. }
%  \label{fig:b235}
%\end{figure}
%
%

\section{Conclusions}

{{Lean hydrogen-air flames freely propagating in 
planar narrow channels with conducting walls were investigated by direct numerical simulation, using  detailed chemical kinetics and transport.
A channel size of $1$ mm, relevant to the micro combustion regime, and
lean mixture conditions with $\phi=0.4$, corresponding to $Le<1$,  were chosen as fixed parameters, while the reactants flow rate and the heat transfer parameter were varied. 

 The simulation  results show that double solutions, symmetric and non-symmetric, can coexist,  which is a confirmation of earlier results obtained within a thermo-difusive (constant density) approximation 
for $Le<1$ flames \cite{KurdyumovJimenez2014}.
Even if no stability analysis has been performed, we have verified that 
when the two solutions exist, the symmetric flame solution is unstable 
to small perturbations and the non-symmetric flame is
stable. This agrees with our previous stability analysis results \cite{Kurdyumov2011, KurdyumovJimenez2014}.

{Our study confirms as well the findings of \cite{KurdyumovJimenez2014}, where a gap was obtained in the flame response curve, indicating that
heat losses can make flame propagation impossible for small flow rates.}
{In addition, because of the adopted DNS approach,} it provides a quantitative estimation of the range of 
conditions where  this flame quenching by heat losses through the wall can be expected. 
The more vigorous burning associated to non-symmetric flames, with combustion being up to five times faster than in symmetric flames, 
 results in self-sustained combustion for a larger parametric range.  
%Another finding of practical  relevance is that the critical flow rate for flashback, corresponding to non-symmetric flames is very different (and much larger) to that predicted for symmetric solutions.
This has important safety implications, as
numerical estimations of the risk of flashback  in small-size conducts containing lean hydrogen mixtures would  erroneously predict lower flashback risk if symmetric conditions are imposed.
 Of course the same is potentially true for any $Le<1$ reacting mixture.
%%erroneous estimations Given that previous 
%estimations of flashback conditions have
%been conducted assuming symmetry of the flame, the present results call 
%for a review of the estimations of the range of conditions where 
%flashback can occur in small-size conducts with lean hydrogen mixtures.

The present study confirms that CFD calculations used in the design of small size combustion devices should avoid the {common} practice of simplifiying the geometry by assuming that the symmetry of the cold flow is conserved. The possibility of symmetry breaking bifurcations needs to be taken into account to reliably predict the
existence and stability  range of flames in small devices.

%
%Finally, our study shows that 
%when the Soret effect is included non-symmetric and symmetric solutions are also found, and that the break of symmetry appears at roughly the same flow rate value.
%The main effect of thermal diffusion appears to be  an important increase in the burning rate at large flow rates, 
% corresponding to enhanced transport of hydrogen to the reaction zone by the  high temperature gradients
% induced by large curvatures and superadiabatic temperatures.}}
%

\section*{Acknowledgments}
This work was supported by grants  \#CSD2010-00011 (MINECO-CONSOLIDER) and \#ENE2015-65852-C2-2-R (MINECO/FEDER), as well as by the EU H2020 Programme and MCTI/RNP-Brazil under the HPC4E Project, grant agreemnt no 689772.

\section*{References}
\bibliographystyle{elsarticle/model1a-num-names}
\bibliography{channelflames}

\clearpage

\end{document}